\documentclass[12pt,twoside]{article}   
\usepackage{amsmath, amssymb}
\usepackage{multirow}
\usepackage{booktabs}
\usepackage[super,sort&compress,comma]{natbib}

\usepackage{fancyhdr}		

\usepackage[section]{placeins}   %

\usepackage[symbol]{footmisc}
\usepackage{graphicx}

\makeatletter \renewcommand\@biblabel[1]{$^{#1}$} \makeatother
\newcommand{\cen}[1]{\begin{center} #1 \end{center}}

\usepackage[mathlines]{lineno}

\usepackage{hyperref}
\hypersetup{ colorlinks,
    citecolor=blue,
    filecolor=blue,
    linkcolor=blue,
    urlcolor=blue
}

\usepackage[symbol]{footmisc}

\usepackage{xcolor}

\definecolor{gray}{rgb}{0.6,0.6,0.6}
\definecolor{red}{rgb}{0.85,0,0}
\definecolor{green}{rgb}{0,0.85,0}
\definecolor{blue}{rgb}{0,0,0.85}
\definecolor{beige}{rgb}{0.92,0.87,0.78}
\usepackage[all]{hypcap}    
\begin{document}

\cen{\sf {\Large {\bfseries Report on the AAPM Grand Challenge on deep generative modeling for learning medical image statistics} \\  
\vspace*{10mm}

Rucha Deshpande$^{*,1}$, Varun A. Kelkar\footnote[1]{Equal contribution.}$^{,2,}$\footnote[2]{Currently at Analog Devices, Inc., Boston, Massachusetts, USA.}, Dimitrios Gotsis$^{2}$, Prabhat Kc $^{3}$, Rongping Zeng$^{3}$, Kyle J. Myers$^{4}$, Frank J. Brooks$^{5,6}$,\\ and Mark A. Anastasio$^{2,5,6}$
} \\
\vspace{2mm}

$^{1}$Dept. of Biomedical Engineering, Washington University in St. Louis, St. Louis, Missouri, USA\\
$^{2}$Dept. of Electrical and Computer Engineering, University of Illinois Urbana-Champaign, Urbana, Illinois, USA\\
$^{3}$Center for Devices and Radiological Health, Food and Drug Administration, Silver Spring, Maryland, USA\\
$^{4}$Puente Solutions, LLC, Phoenix, Arizona, USA \\
$^{5}$Center for Label-free Imaging and Multiscale Biophotonics (CLIMB), University of Illinois Urbana-Champaign, Urbana, Illinois, USA\\
$^{6}$Dept. of Bioengineering, University of Illinois at Urbana-Champaign, Urbana, Illinois, USA\\

\vspace{5mm}
}

\pagenumbering{roman}
\setcounter{page}{1}
\pagestyle{plain}
Author to whom correspondence should be addressed: Mark A. Anastasio. Email: maa@illinois.edu \\

\begin{abstract}
\noindent {\bf Background:} The findings of the 2023 AAPM Grand Challenge on Deep Generative Modeling for Learning Medical Image Statistics are reported in this Special Report.\\ 
{\bf Purpose:} The goal of this challenge was to promote the development of deep generative models for medical imaging and to emphasize the need for their  domain-relevant assessments via the analysis of relevant image statistics.\\
{\bf Methods:} As part of this Grand Challenge, a common training dataset and an evaluation procedure was developed for benchmarking deep generative models for medical image synthesis. To create the training dataset, an established 3D virtual breast phantom was adapted.
The resulting dataset comprised about 108,000 images of size 512$\times$512. For the evaluation of submissions to the Challenge, an ensemble of 10,000 DGM-generated images from each submission was employed. The evaluation procedure consisted of two stages. In the first stage, a preliminary check for memorization and image quality (via the Fr\'echet Inception Distance (FID)) was performed. Submissions that passed the first stage were then evaluated for the reproducibility of image statistics corresponding to several feature families including texture, morphology, image moments, fractal statistics and skeleton statistics. A summary measure in this feature space was employed to rank the submissions. Additional analyses of submissions was performed to assess DGM performance specific to individual feature families, the four classes in the training data, and also to identify various artifacts.\\
{\bf Results:} Fifty-eight submissions from 12 unique users were received for this Challenge. Out of these 12 submissions, 9 submissions passed the first stage of evaluation and were eligible for ranking. The top-ranked submission employed a conditional latent diffusion model, whereas the joint runners-up employed a generative adversarial network, followed by another network for image superresolution. In general, we observed that the overall ranking of the top 9 submissions according to our evaluation method (i) did not match the FID-based ranking, and (ii) differed with respect to individual feature families. Another important finding from our additional analyses was that different DGMs demonstrated similar kinds of artifacts.\\
{\bf Conclusions:} This Grand Challenge highlighted the need for domain-specific evaluation to further DGM design as well as deployment. It also demonstrated that the specification of a DGM may differ depending on its intended use.\\

\end{abstract}

\newpage     

\pagenumbering{arabic}
\setcounter{page}{1}
\section{Introduction}
Over the past several years, deep generative models (DGMs), such as generative adversarial networks (GANs) and diffusion models, have gained popularity for their ability to generate perceptually realistic images \cite{foster2019generative, karras2020analyzing, dhariwal2021diffusion}. DGMs are being actively explored for a variety of medical imaging applications \cite{yi2019generative} such as data sharing\cite{dumont2021overcoming}, image restoration \cite{sagheer2020review, kaur2023complete}, image reconstruction \cite{zhang2020review, song2021solving, marinescu2020bayesian}, image translation \cite{kazeminia2020gans, lutnick2020generative}, and objective image quality assessment \cite{zhou2019approximating}. DGMs hold potential for creating large image sets for applications where clinical data are limited, as may be the case when developing an AI/ML algorithm for use on a new imaging modality or an indication for which little clinical data is available. Another important potential use for DGMs is the creation of datasets for use in the task-based evaluation of imaging systems that make use of realistic images and tasks \cite{zhou2022learning}.


While DGMs that include a variety of GANs and, more recently, diffusion-based models, have demonstrated a remarkable ability to synthesize images that possess high perceptual quality, their evaluations remain limited in scope \cite{theis2015note, borji2022pros, bandi2023power, stein2024exposing, okawa2024compositional}. This is especially true within the context of medical imaging applications, where images are typically employed for specific diagnostic tasks.  In most reported studies of DGMs, ensemble-based metrics such as the Fr\'echet Inception distance (FID) or inception score (IS) have been employed as evaluation metrics \cite{o2021pre, woodland2022evaluating, tronchin2021evaluating}. However, it is well-known that DGMs possessing favorable FID or IS scores can still produce images that are degraded by impactful errors and/or can fail to correctly reproduce image statistics that are relevant and important to a medical imaging task \cite{deshpande2021method, bhadra2021hallucinations, cohen2018distribution}. Moreover, distributions of medical images typically comprise multiple classes or modes, and it is known that DGMs may produce significant errors when synthesizing images from a mode that is infrequently encountered during training \cite{kelkar2023assessing}.

The medical imaging community has begun to develop procedures for assessing DGMs with consideration of domain-relevant factors. Kelkar et al.\! \cite{kelkar2023assessing}assessed the ability of a state-of-the-art GAN to learn canonical statistics of several stochastic image models (SIMs) that are relevant to medical imaging applications and investigated the extent to which task-agnostic measures such as FID score compare with clinically meaningful measures of image quality. That study revealed that the GANs considered could consistently produce images of high perceptual quality and were able to accurately and consistently reproduce some basic first and second-order image statistics but could fail to capture other important statistics. It also confirmed that the FID score could not always be used as a proxy for task-relevant measures when tuning training hyperparameters. Lee et al.\! \cite{lee2023impact} explored conditional GANs for simulating mammograms and found that artifacts were pervasive in the simulated images. Despite this, they concluded that the simulated images still provided complementary information for cancer detection when they were combined with real mammograms.

Deshpande et al.\! \cite{deshpande2021method} designed several stochastic context models (SCMs) of distinct image features that could be employed to assess the ability of a DGM to correctly reproduce spatial context. They observed that ensembles of generated images could be visually accurate and possess high accuracy based on ensemble measures but could fail to reproduce the known spatial arrangements of image features. 
In a follow-up study that leveraged SCMs \cite{deshpande2023assessing}, it was found that denoising diffusion probabilistic models hold significant capacity for generating contextually correct images that are `interpolated' between training samples, which may benefit data-augmentation tasks \cite{zhang2023diffusion} in ways that GANs cannot.
While these and other studies \cite{tronchin2021evaluating, scholz2023metrics, muller2023multimodal, jang2022assessment, uzunova2022systematic} have provided valuable insights into the performance of DGMs, there remains a need for more widespread application-relevant assessments of DGMs in the field of medical imaging.

To address this need and promote meaningful assessments and refinements of DGMs for medical imaging applications,
the American Association of Physicists in Medicine (AAPM) hosted the Deep Generative Modeling for Learning Medical Image Statistics Challenge, or the DGM-Image Statistics Challenge for short. Each year the AAPM issues a call for Grand Challenges in order ``to assess or improve the use of medical imaging in both diagnostic and therapeutic applications.” The DGM-Image Statistics Challenge invited participants to develop or refine generative models that can accurately reproduce image statistics that are important and relevant to medical imaging applications, including the evaluation of imaging systems as well as for use in the training and testing of AI/ML algorithms. Through the DGM-Image Statistics Challenge, we provided a dataset, a standardized evaluation procedure, and a benchmark for evaluating future generative models for medical image synthesis. 
 A description of the DGM-Image Statistics Challenge framework and a reporting and discussion of the results are provided in this manuscript.

\section{DGM-Image Statistics Challenge Overview}

\begin{figure}[hbt]
\centering
\includegraphics[width=\linewidth]{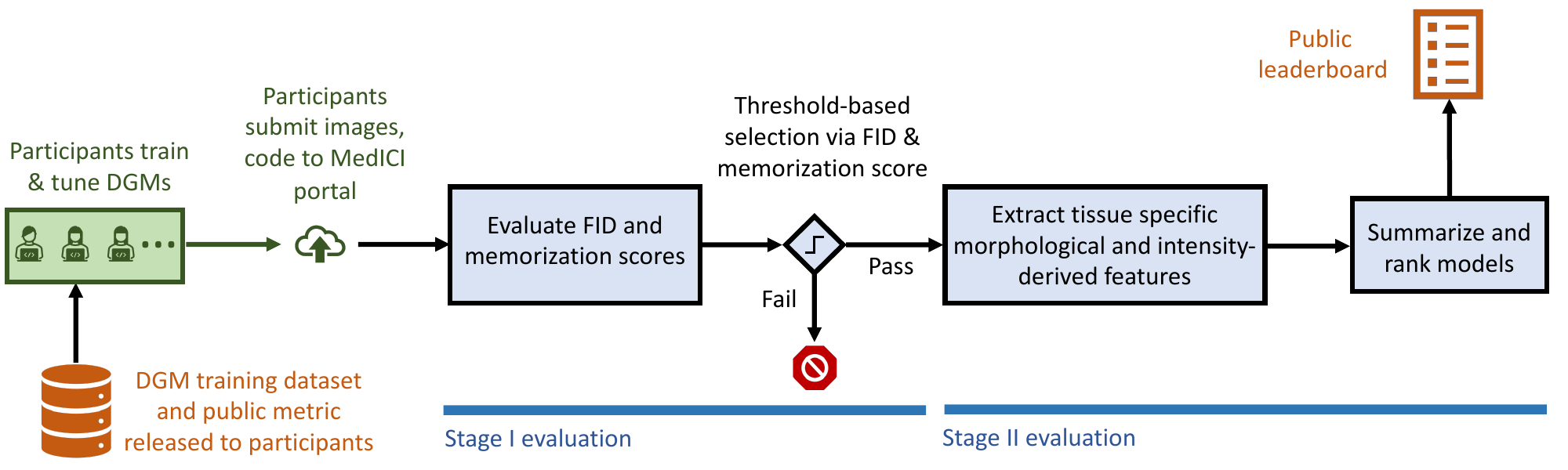} 
\caption{The DGM challenge workflow.}
\label{fig:dgm_schematic}
\end{figure}

The broad objectives of the DGM-Image Statistics Challenge were to 1) facilitate the development and refinement of DGMs that can reproduce several key image statistics that are known to be useful for the specified medical image application, and 2) promote the use of meaningful DGM assessment procedures.
Specifically, the DGM-Image Statistics Challenge sought to identify the DGM learned from a provided data set that most accurately reproduces the distribution of certain morphological and intensity-derived statistics in the provided data as well as other relevant features described below, while still producing perceptually realistic images and avoiding overfitting/memorization of the training data. A summary metric was derived from these statistics and used to score and rank participants in order to declare a winner and a runner-up. A schematic illustrating the workflow of the challenge is shown in \autoref{fig:dgm_schematic}.

A unique element of the DGM-Image Statistics challenge was its focus on the development and evaluation of DGMs for creating ensembles of digital objects that could serve as inputs to a simulated medical imaging system. These so-called Stochastic Object Models (SOMs) enable the evaluation of imaging systems through in silico clinical trials, facilitating the assessment and optimization of medical imaging systems before they are built \cite{badano2018evaluation}. In silico trials offer an important alternative to physical clinical trials for the assessment of novel medical imaging technologies \cite{abadi2020virtual}. Models for inputs to the imaging system, as well as for the imaging system itself and the image interpretation process, are needed for the execution of an in silico clinical trial. The DGM-Image Statistics Challenge probed the degree to which suitable SOMs might be synthesized using generative methods.

The provided training dataset comprised coronal slices from a series of anthropomorphic breast phantoms from the VICTRE project \cite{sharma2019silico,badano2017silico}, as described in the Methods section. The rationale for this training dataset selection was twofold: 1) The VICTRE phantoms were used in an in silico trial that successfully replicated an actual physical trial. Thus a demonstration that a DGM could synthesize the salient features of these objects would provide some confidence that such models could be used to synthesize SOMs in the future for other purposes; 2)  Because the VICTRE phantoms are procedurally generated and are not degraded by an imaging process, their statistical properties are known. They could therefore be readily compared to the statistics of data obtained by the competing DGMs submitted by the Challenge participants.

\section{Methods}
\subsection{Methods: Challenge Logistics}\label{sec:challenge_logistics}

The challenge was divided into two phases. In phase 1, the participants were required to submit 10,000 independent and identically distributed images sampled from the generative model trained by them on the provided training data, along with a one-page description of their approach. In phase 2, the participants were additionally required to submit a code implementation of their approach, containerized using Docker \cite{docker}. Participants were only allowed to use the provided training data 
to develop their models. Furthermore, the participants were limited to DGMs that were capable of generating 10,000 images within 12 hours on a single NVIDIA V100 graphics processing unit (GPU) with 16 GB RAM.  

During both the phases, the participants' submissions were evaluated via a two-stage evaluation pipeline. 
The first stage of the evaluation, based on the Fr\'echet Inception Distance (FID) score and a custom memorization metric, was used to rule out submissions consisting of images having obviously poor perceptual quality, or those that memorized the images from the training data. 
Submissions were restricted to DGMs that were estimated from the provided training data. In the second stage of the evaluation, a figure of merit based on more than 3,000 textural, morphological and intensity-derived features extracted from the images was used to rank the participants. The specific features underlying the figure of merit were not disclosed to the participants. Instead, the procedure and the software implementation for computing a simplified measure based on a smaller subset of the identified features was made available to the participants via the challenge portal in order to provide feedback regarding their models. Details of the evaluation pipeline are presented in \autoref{sec:evaluation_strategy}
In phase 2, the code provided by the participants was used to generate 10,000 images on which the two-stage evaluation was performed. Additionally, the code and the images from the best submission from each of the participants in phase 2 was manually validated by one of the organizers. 

The challenge was hosted on the Medical Imaging Challenge Infrastructure (MedICI) platform \cite{medici}.
The scores of the participants were made public on a leaderboard hosted on the Challenge platform\footnote[3]{The link to the challenge leaderboard is as follows: \url{https://www.aapm.org/GrandChallenge/DGM-Image/dgm-leaderboard.pdf}}. Scores of a baseline StyleGAN2 model trained by the organizers on the provided data were also displayed on the leaderboard. For the received submissions, all computation pertinent to the evaluation pipeline was performed on an Intel Xeon v4 CPU and, whenever needed, a 16 GB NVIDIA V100 GPU hosted on the CoreWeave cloud computing platform \cite{coreweave}. \autoref{tab:timeline} shows the timeline for the challenge.

\begin{table}[]\label{tab:timeline}
\caption{Timeline for the DGM-Image challenge}
\vspace{5pt}
\begin{tabular}{@{}ll@{}}
\toprule
January 13, 2023 & Challenge announced, participant registration opened                         \\
January 16, 2023 & Training set released                                                        \\
May 19, 2023     & Deadline for Phase 1 submission                                              \\
May 29, 2023     & Deadline for Phase 2 submission                                              \\
June 14, 2023    & Participants and top-ranked finishers contacted with results                 \\
July 27, 2023    & Top two finishers presented their results at the AAPM annual meeting \\ \bottomrule
\end{tabular}
\vspace{10pt}
\end{table}

\subsection{Methods: Data design}

Although anatomical realism was important in the design of training data for the challenge, some practical aspects such as the typical computational requirements of training DGMs on large images, as well as robustness of post-hoc evaluation were also considered. A previously published stochastic object model of the human female breast: VICTRE \cite{badano2018evaluation}, was adapted for this grand challenge. From each 3D volume generated at a voxel resolution of 0.1 mm${^3}$ via the VICTRE tool, fifteen equidistant 2D slices were extracted from the central third of the volume. Eight different tissues or structures were described by VICTRE within this sub-volume. Only 4 of those tissue types: skin, fat, glandular tissue, and ligaments were retained in the adapted version for the challenge. This choice of tissues was based on (i) the presence of all 4 tissue types in each 2D slice, (ii) generally distinct tissue properties for the chosen imaging modality: x-rays at 30 keV, and (iii) the structural variety provided by the tissue types. Thus, this choice of tissues contributed to the robustness and utility of the evaluation method.

Structures belonging to the four omitted tissue types (artery, vein, duct, and terminal duct lobular units) were replaced by glandular tissue, which was the most similar tissue in terms of attenuation co-efficients. Each slice was then downsampled to size 512$\times$512. Even with this downsampling, the thin ligament structures in the original slices were retained. The data dimensions were chosen based on compute requirements of training modern DGMs on large images \cite{karras2020analyzing, dhariwal2021diffusion, sauer2022stylegan}, and the time window of the challenge. The downsampling process involved the following: (i) the breast region in an image was identified with a bounding box, (ii) the four chosen tissue types within the bounding box were separated into four distinct binary arrays such that each pixel location was foreground only for one tissue, (iii) for each of these arrays, the coordinates of the foreground pixels were transformed to match a 512$\times$512 array with a centered breast region, (iv) the ligament array was skeletonized \cite{van2014scikit}, (v) all tissue arrays were thresholded, and (vi) the resulting four binary arrays –-- with exclusive tissue identity at each pixel location --- were combined to yield a single image. Next, tissue specific-intensity distributions were pre-defined such that the relative tissue attenuation properties were largely maintained in the grayscale intensity range of 0 to 255. An exception was made for ligaments and skin tissues, which had very similar attenuation properties. The grayscale distributions for these two tissues were slightly adjusted to enhance their separability and aid post-hoc analyses. The final four tissue-specific intensity distributions were specified as distinct Beta distributions: 
\begin{equation}
    \begin{aligned}
    \mathrm{t_{fat}}\sim 60\:X + 52,\; \mathrm{where}\; X \sim \mathrm{Beta}(\alpha=2,\beta=4), \\
    \mathrm{t_{glandular}}\sim 96\:X + 128,\; \mathrm{where}\; X \sim \mathrm{Beta}(\alpha=4,\beta=2), \\
    \mathrm{t_{skin}}\sim 16\:X + 228,\; \mathrm{where}\; X \sim \mathrm{Beta}(\alpha=3,\beta=3), \\ 
    \mathrm{t_{ligaments}}\sim 16\:X + 232,\; \mathrm{where}\; X \sim \mathrm{Beta}(\alpha=3,\beta=3). 
    \end{aligned}
    \label{eq:intensities}
\end{equation}

 Next, variates from these intensity distributions were assigned to appropriate tissue locations as follows. Four arrays of size 512$\times$512 were generated from each distribution for a single 2D slice. A texture was imposed on each array via a Gaussian filter with smoothing parameter $\sigma=0.8$, similar to the process described in \cite{li20213}. This resulted in slightly correlated pixels, thus, generating a prescribed texture, which could be then tested as part of the evaluation framework. Note that the extent of Gaussian smoothing was chosen subjectively, and not to a known correlation in tissue attenuation. After Gaussian smoothing, the histograms of the resulting arrays were transformed to ensure that the prescribed intensity distributions were maintained. Each of the four arrays was then masked according to the known tissue locations and combined to yield the final 2D image. 

\begin{figure}[hbt]
\centering
\includegraphics[width=0.4\linewidth]{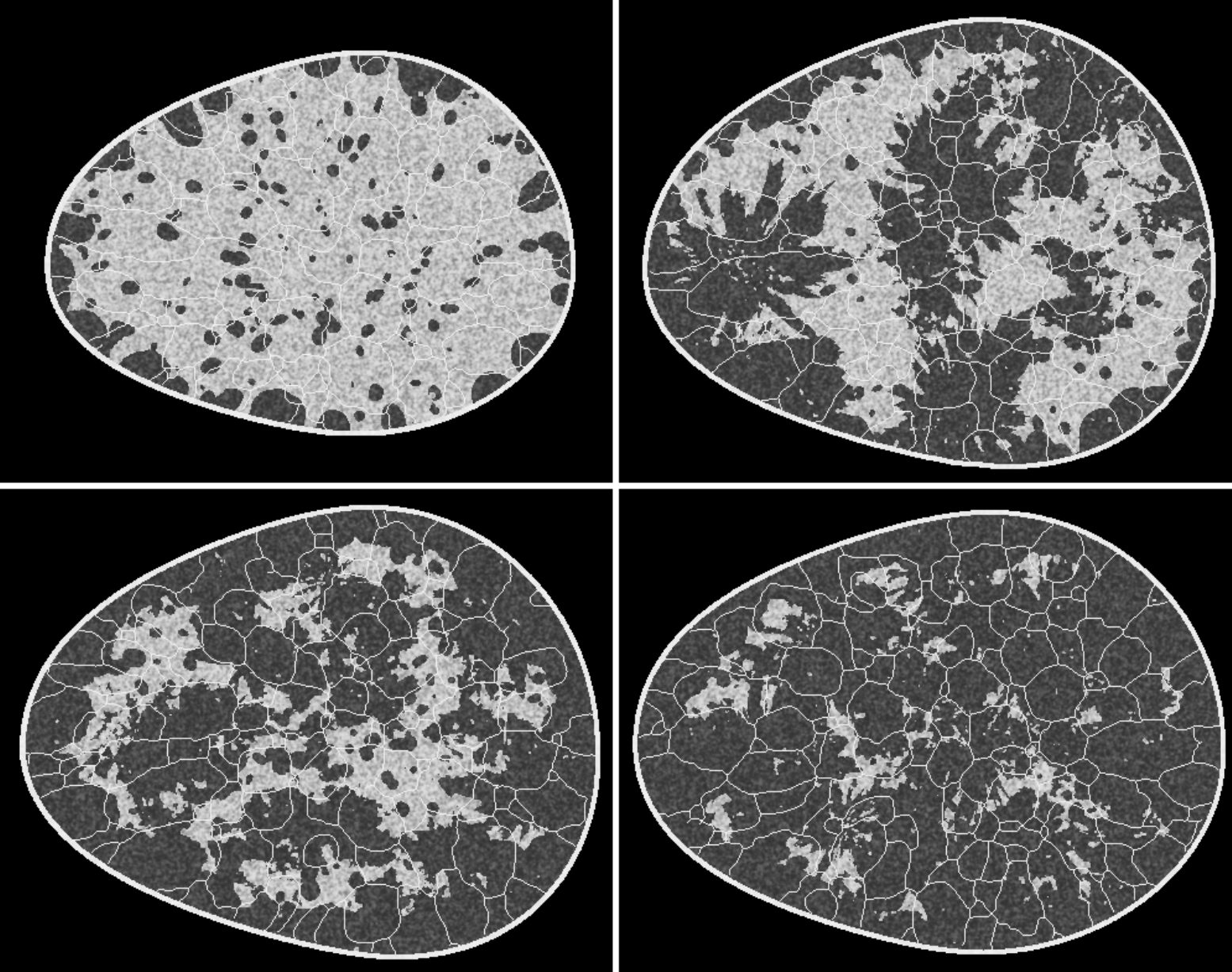} 
\caption{Sample images from the training dataset corresponding to four classes: dense (upper left), heterogeneous (upper right), scattered (lower left), and fatty (lower right). Class information was not provided explicitly to the participants.}
\label{fig:reals}
\end{figure}

The training ensemble consisted of about 108,000, 8-bit images of size $512\times512$, saved via lossless compression, and was made available to the participants after registration for the challenge. This training ensemble comprised four breast types, as determined by the Breast Imaging Reporting and Data System (BI-RADS) classification system \cite{liberman2002breast}. In accordance with population prevalence \cite{liberman2002breast}, the four breast types, namely, fatty, scattered, heterogeneous and dense, were represented in the ratio of 1:4:4:1 within the training dataset. However, this information was not explicitly provided to the participants during the challenge. The challenge dataset is now publicly available along with the breast type label for each image \cite{challenge_data}. Sample images from each of the four classes in the challenge dataset are shown in \autoref{fig:reals}.

\subsection{Methods: Evaluation strategy}\label{sec:evaluation_strategy}
From each DGM submission, an ensemble of 10,000 images was generated for evaluation. As described in \autoref{sec:challenge_logistics}, the first stage of evaluation identified entries eligible for ranking via FID scores and a memorization measure. The memorization measure was the cross-correlation of binary masks representing the fatty-glandular tissue boundary, computed for all images in a DGM-generated ensemble against all images in the training ensemble. The boundaries between the two tissues were obtained via a series of simple filtering operations followed by thresholding. An image in a generated ensemble was marked as memorized if the memorization measure exceeded a value of 0.9 on a [0, 1] scale. This threshold was determined as one standard deviation greater than the maximum value observed after calibrating this measure on a subset of 3000 randomly chosen images from the training ensemble against the remainder of the ensemble (about 108,000 images). 

For the second stage of evaluation, all images in the training and DGM-generated ensembles were first segmented to obtain individual tissue regions. The segmentation process involved global thresholding that employed knowledge of the prescribed tissue-specific intensity distributions. Specifically, the intensity ranges employed for thresholding were: fatty tissue [30, 120], glandular tissue [120, 226], and skin/ligaments [226, 255]. Note that intensity values below 30 were allocated as background or the absence of any tissue. In addition, although skin and ligaments have highly similar intensity distributions, they are easily distinguishable based on their location, i.e., peripheral (skin) or central (ligaments). Thus, a single image yielded the following tissue labels after segmentation: fatty tissue (F), glandular tissue (G), ligaments (L), skin (S). Features were then extracted from individual tissues (F,G,L,S) or the full breast slice (B), as indicated in the parentheses that follow the feature descriptor as follows: (i) texture features (B) \cite{castella2008mammographic, kelkar2023assessing}, (ii) morphological features (F, G, B) \cite{van2014scikit}, (iii) skeleton statistics (L) \cite{nunez2018new}, (iv) so-called ``fractal" features – box dimension and box-counting lacunarity (F, G, L), consistent with their popular formulations \cite{falconer2004fractal, smith1996fractal}, (v) moments – raw, central, normalized, Hu, and their weighted versions (F, G, B) \cite{van2014scikit}, and (vi) ratio of fatty to glandular tissue (B) \cite{liberman2002breast, kelkar2023assessing}. For the computation of texture features, all data were binned to 64 gray levels, which were determined to be sufficient via the Freedman-Diaconis rule \cite{freedman1981histogram} for the training dataset. Note that although the texture features were computed via our in-house implementation, their formulations are well-documented \cite{van2017computational}. The feature sets above were chosen because they have been extensively employed for image classification and object recognition via conventional methods \cite{chaudhuri1995texture, prokop1992survey, rogowska2000overview, flusser2009moments, nunez2018new}, and we do not claim that these features are sufficient to describe diagnostic aspects of biomedical images.

Features that yielded multiple values for a single image, e.g., area of each disconnected fatty tissue region in an image, were summarized as: total count, mean, standard deviation, minimum value, maximum value and quartiles for each realization. In all, over 3400 popular features were extracted from each slice over all feature families. Principal component analysis was performed on all features corresponding to the training data and each DGM-generated ensemble was projected into this principal component (PC) space after feature extraction. To obtain a baseline distribution corresponding to the training data, two data points represented as 10D vectors in the top-10 PC space were chosen at random and the cosine distance between them was computed; this process was repeated 10,000 times. For the DGM-generated ensemble, a similar computation was performed for one data point from the training ensemble and another from the DGM-generated ensemble, both represented as 10D vectors in the PC space of the training data. The two resulting distributions of cosine distances were then compared via the Kolmogorov-Smirnoff (KS) \cite{chakravarti1967handbook} statistic. The procedure was repeated 1000 times on bootstrapped datasets to estimate the uncertainty in the KS statistic, and the resulting mean value of the KS statistic was employed to determine the final rankings in the challenge. 

The evaluation pipeline described above was also employed for computing the public metric, and differed only in the choice of features. Only nine features derived from the intensity histogram and tissue areas were employed in the public metric computation.
For additional class-based analyses, not part of the ranking framework, a four-class classifier with a VGG-16 \cite{simonyan2014very} backbone was trained on 5000 images per-class, for 400 epochs. Recall that the four breast classes correspond to dense, heterogeneous, scattered, and fatty breast types. A validation set of 1500 images per-class was employed, and the model with the least validation loss was selected for inference. The training and validation datasets were distinct from the public dataset for the challenge. Calibration of the classifier on 3000 images per-class from the challenge dataset showed error-rates of 0\%, 0.07\%, 0.37\%, 0.87\% for the four classes. This classifier was employed on all images from the final submissions to predict a single class label for each image.

\subsection{Methods: Participants' Methods} 

In this section, we summarize the methods used by the participants and broadly describe the various strategies developed by them. This information was based on the one-page report submitted by the participants. The groups that did not submit a report are excluded from this survey.

All the participants employed existing state-of-the-art generative modeling approaches as a starting point, and developed additional strategies to improve the performance of their models on the provided training data. The methods were split between those using advanced GAN-based approaches, and diffusion models-based approaches. The groups that developed GAN-based approaches included \textit{K7}, \textit{A8}, \textit{V4}, \textit{S4}, \textit{J5}, and \textit{H1}. 
The groups that developed approaches based on diffusion models included \textit{D9}, \textit{C2}, and \textit{M3}. More specifically, \textit{D9} and \textit{M3} used conditional latent diffusion models \cite{latent_diffusion, medfusion}, whereas \textit{C2} used a denoising diffusion GAN \cite{ddg}.

Some groups, such as \textit{S4}, \textit{J5} and \textit{V4} conducted an extensive hyperparameter search with GAN models, (such as those from the StyleGAN family), without significant modifications the underlying architecture. Their results revealed a large performance gap between models with different hyperparameter configurations, which underscored the significant improvements possible by proper hyperparameter selection. Some groups, such as \textit{S4}, \textit{J5}, \textit{V4}, and \textit{H1} experimented with an adaptive discriminator augmentation (ADA) strategies first proposed by Karras, \textit{et al.} \cite{stylegan2_ada} These included augmentation of the inputs to the discriminator using, for instance, pixel blitting, geometric transformations, and color transformations. However, not all participants found ADA to be helpful for improving the performance of the GAN. 

Several groups utilized additional processing of the images generated by the base GAN or diffusion models to improve their quality. For instance, \textit{D9} and \textit{H1} used classical image processing techniques such as thresholding and filtering. \textit{K7}, \textit{A8}, and \textit{C2} performed deep learning-based image superresolution of the outputs of the generative model. Thus, even if a generative model could only reliably generate coarse-scaled image features, it was still useful, and the fine-scaled features could then be added via a second model for image superresolution.

An interesting characteristic of the top three approaches was that they all relied on conditional image generation using conditioning inputs that implicitly incorporated textural or breast type-relevant information. For example, \textit{D9} developed a conditional latent diffusion model, with a bivariate conditioning input of the area of the breast phantom, and the fat-to-glandular ratio. On the other hand, \textit{K7} and \textit{A8} first segregated the provided data into four classes using \textit{k}-nearest neighbor (KNN) clustering based on the provided public metric features and additional features extracted from breast density classifiers that were trained on screening mammography exams \cite{breast_density_classifier}. A class-conditional generative model with these class labels as the conditioning inputs was trained subsequently. 

\section{Results}

\subsection{Participation summary}
The challenge received 58 submissions from 12 unique usernames. Out of the 12 unique submissions, six were from teams in the United States, two from India, and one each from Belgium, Brazil, Canada, and China. Split by sector, seven submissions came from academia or non-governmental organizations, two from the industry, and three from independent contributors or contributors with unknown affiliations. 

\subsection{Results: Overall Results}
\begin{figure}[h!bt]
\vspace{10pt}
\centering
\includegraphics[width=0.8\linewidth]{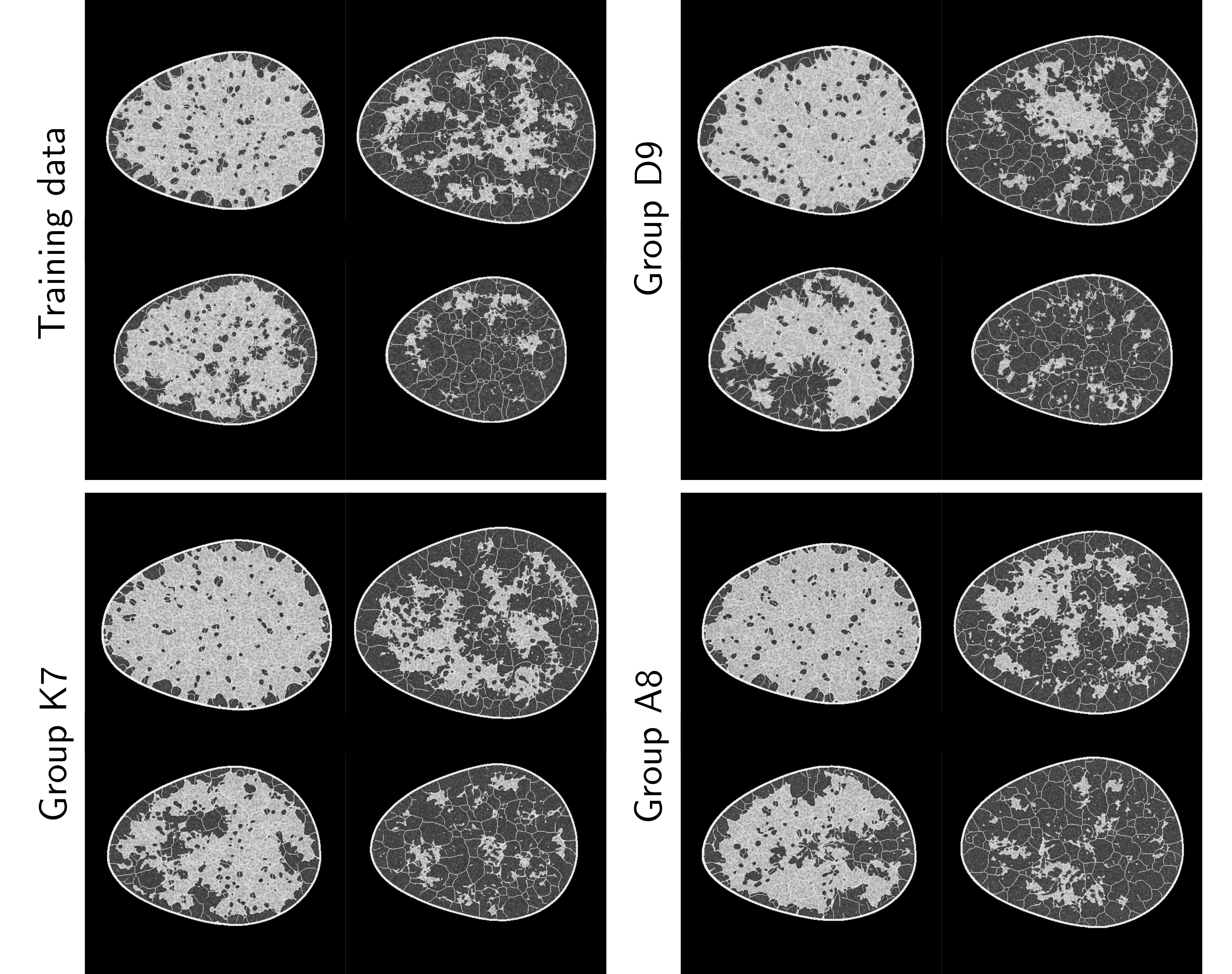} 
\caption{Images generated by the top three approaches alongside the images from the training data.}
\label{fig:top3_images}
\vspace{10pt}
\end{figure}

\begin{figure}[h!bt]
\vspace{10pt}
\centering
\includegraphics[width=\linewidth]{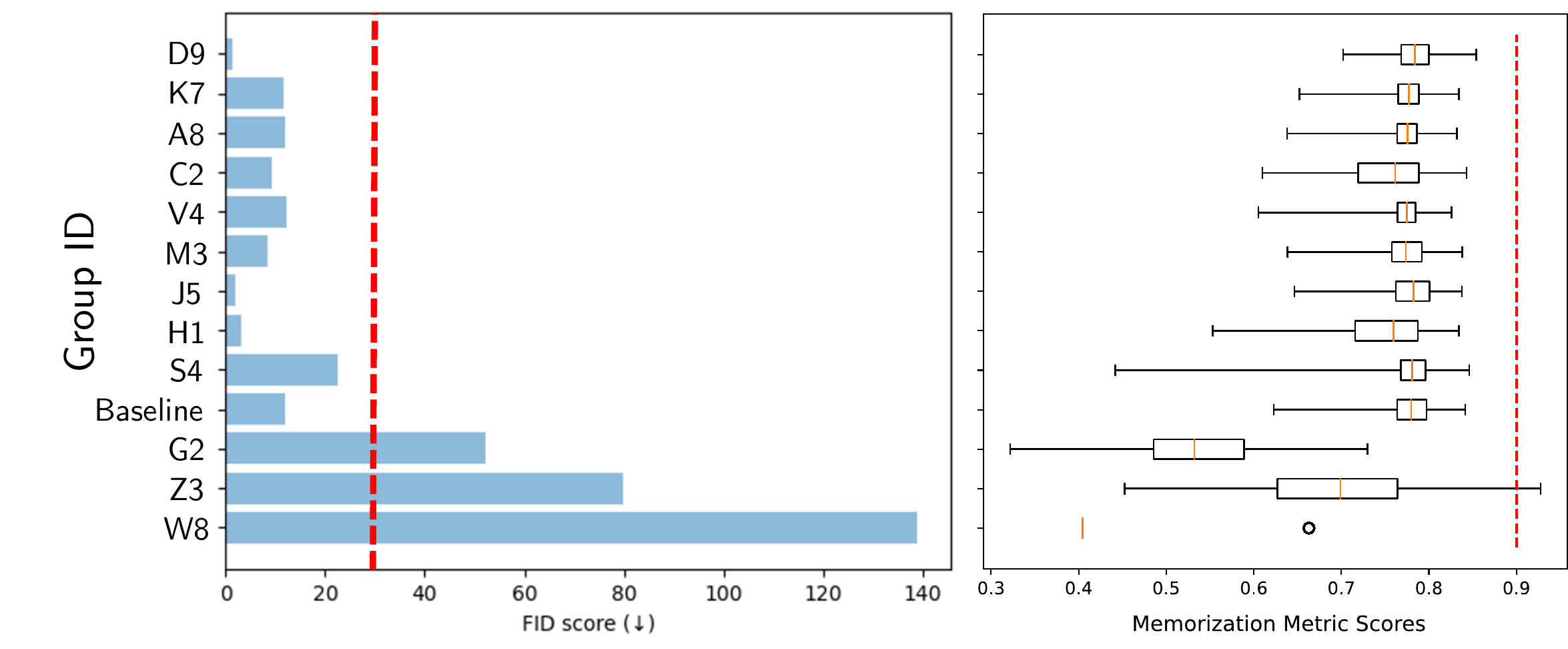} 
\caption{FID and memorization metric scores for the submissions alongside the FID and memorization metric scores of a baseline StyleGAN2 model trained in-house}
\label{fig:fid}
\vspace{10pt}
\end{figure}

\begin{figure}[h!bt]
\vspace{10pt}
\centering
\includegraphics[width=0.7\linewidth]{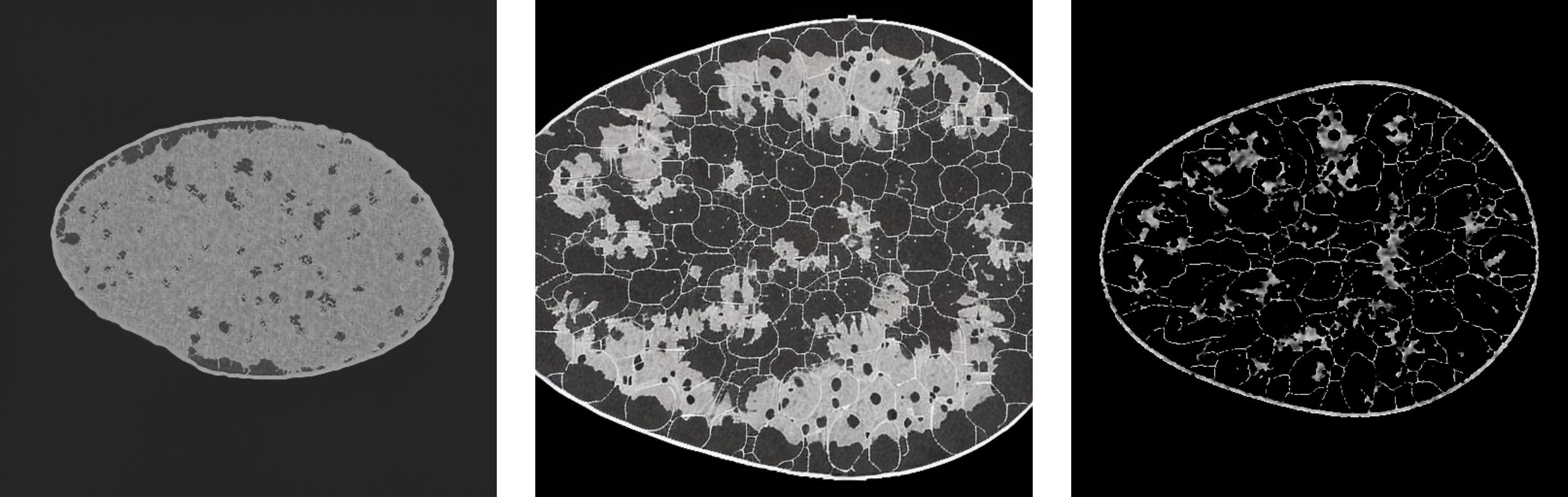} 
\caption{An image each from three submissions that were ruled out in the first stage of evaluation. The images in the left and center positions correspond to the submissions that did not pass the FID threshold, whereas the rightmost image corresponds to the submission that did not pass both the FID and the memorization thresholds.}
\label{fig:bad_images}
\vspace{10pt}
\end{figure}

\autoref{fig:top3_images} shows the images generated by the top three approaches. It can be seen that they display a high degree of perceptual similarity to the provided training images. However, they still retain imperfections, as identified and discussed further in \autoref{sec:further_analysis} The FID scores for the 12 unique submissions are reported in \autoref{fig:fid}a. Boxplots of memorization measure values computed for 3,000 synthetic images with respect to the training data are shown in \autoref{fig:fid}b. The submissions that produced images with memorization measure value $> 0.9$ were ruled to contain memorized images. These images were manually checked to ensure that the images indeed closely resembled specific images from the training dataset, for instance, in terms of the exact placement of the tissue boundaries. Submissions that produced memorized images were ruled out in the first stage of the analysis. An FID score threshold of 30 was chosen so that the submissions that generated obviously visually unrealistic images were also ruled out in the first stage of the analysis. Two submissions failed the FID score check, whereas one submission failed both the FID score and memorization metric check.  An example image from each of these three submissions is shown in \autoref{fig:bad_images}. These three submissions were ruled out in the first stage of the evaluation.

\begin{figure}[h!bt]
\vspace{10pt}
\centering
\includegraphics[width=\linewidth]{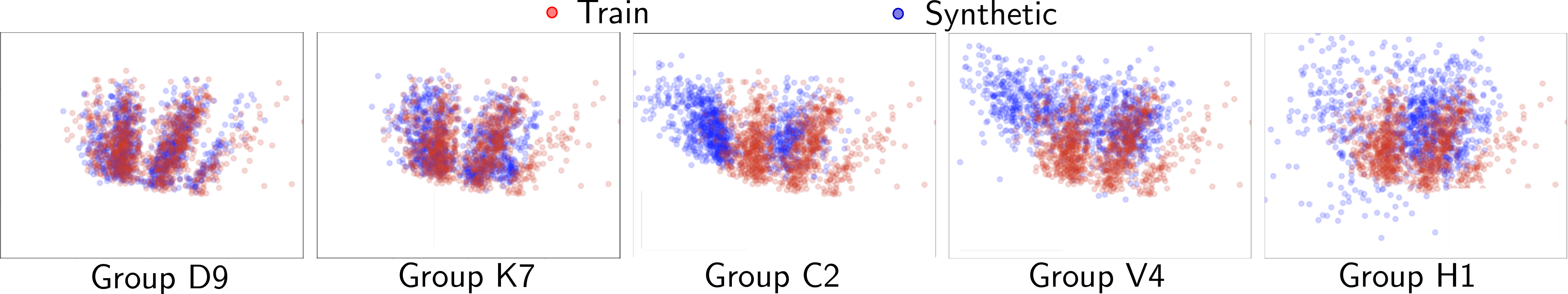} 
\caption{First two principal components of the features extracted from images from submissions ranked 1, 2, 4, 5 and 8.}
\label{fig:pca}
\vspace{10pt}
\end{figure}

The nine submissions that qualified for the second stage of the evaluation were analyzed using the evaluation pipeline described in \autoref{sec:evaluation_strategy} \autoref{fig:pca} shows scatter plots of the first two principal components of the features extracted from DGM-generated images for five out of the final nine submissions. These plots clearly suggest an increasing disparity with rank between the true and the learned empirical PDFs over the top two principal components of the feature data.  

\begin{figure}[h!bt]
\vspace{10pt}
\centering
\includegraphics[width=\linewidth]{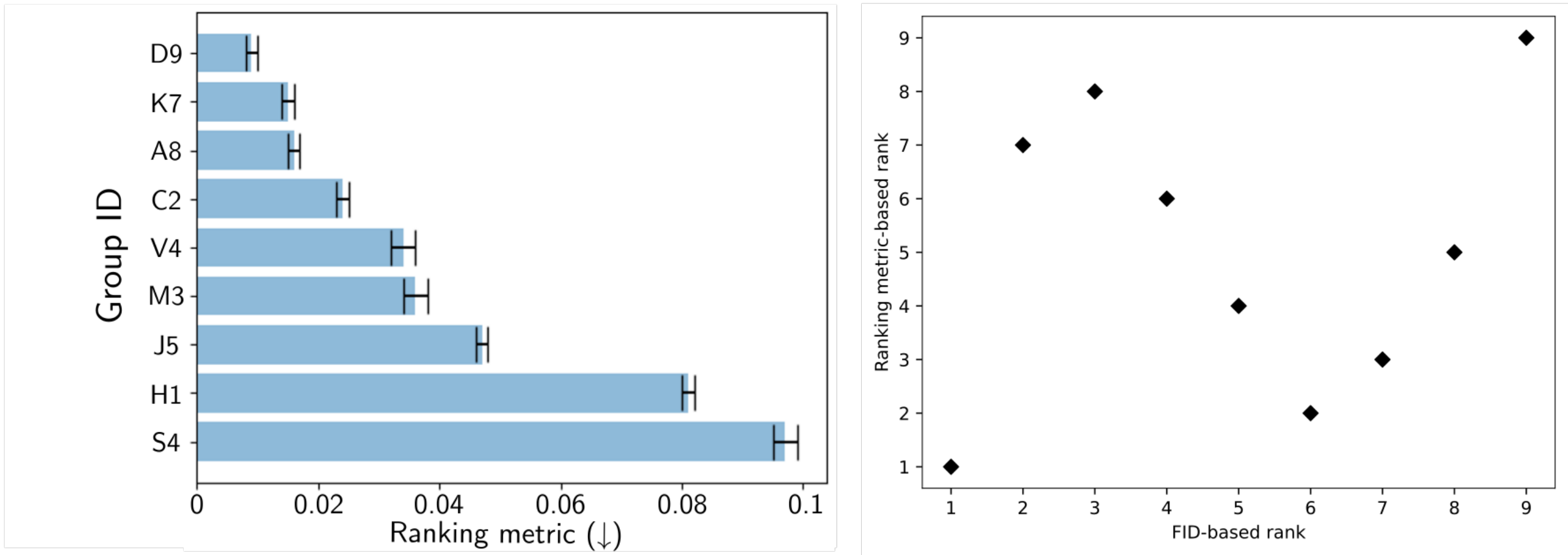} 
\caption{(Left) A histogram showing the ranking metric values for the 9 submissions that passed the FID-based threshold. (Right) Rank of the submissions with respect to FID plotted against the rank of the submissions with respect to the ranking metric. Note that this plot only shows the submissions that have passed the FID-based threshold.}
\label{fig:ranking_metric}
\vspace{10pt}
\end{figure}

\autoref{fig:ranking_metric}a shows a bar plot of the final ranking metric for the nine submissions, and \autoref{fig:ranking_metric}b shows the relationship of their FID-based rank to their rank based on the final ranking metric. It can be seen that for the submissions that were identified to be below a baseline FID-based threshold (30), the rankings based on the FID and the final ranking metric show poor correlation.

Results from the supplementary analyses of the final nine submissions are reported in the following three sub-sections. A representative subset of features from the original evaluation framework were employed for the additional analyses. Note that these analyses were not employed to determine the overall ranking and are presented only to provide additional insights into the different DGMs considered.

\subsection{Results: Performance on individual feature families}

\begin{table}[htb]\label{tab:familyrankings}
\centering
\caption{Submission rankings based on individual feature families. \\ (F/G: fatty to glandular tissue ratio, Moment: normalized image moments)}
\vspace{5pt}
\begin{tabular}{@{}llllllll@{}}
\toprule
Group ID & Overall & Texture & F/G & Moment & Morphology & Fractal & Skeleton \\
D9 & \textbf{1} & 8 & 3 & \textbf{1} & \textbf{1} & 3 & \textbf{1} \\
K7 & 2 & \textbf{1} & 8 & 4 & 5 & \textbf{1} & 3 \\
A8 & 3 & 2 & 4 & 3 & 6 & 2 & 2 \\
C2 & 4 & 4 & \textbf{1} & 2 & 8 & 8 & 4 \\
V4 & 5 & 7 & 2 & 7 & 7 & 9 & 9 \\
M3 & 6 & 9 & 9 & 6 & 3 & 7 & 6 \\
J5 & 7 & 3 & 6 & 5 & 2 & 4 & 7 \\
H1 & 8 & 5 & 5 & 8 & 4 & 5 & 8 \\
S4 & 9 & 6 & 7 & 9 & 9 & 6 & 5 \\

\bottomrule
\end{tabular}
\vspace{10pt}
\end{table}

As described previously, to determine the overall ranking for the Challenge, all features were weighted equally. Rankings for \emph{individual} feature-families are reported in \autoref{tab:familyrankings}. The overall top 3 submissions also ranked between 1 and 3 for several individual feature-families. The best submission performed remarkably well on most feature-families, except on the texture features. On the other hand, some lower ranked submissions were ranked high for a single feature-family (e.g., \textit{J5} on morphological features, \textit{C2} on F/G ratio). Thus, the choice of the ``best submission" may vary based on the image statistics that are deemed important to a specified diagnostic task.  

\subsection{Results: Class-based analyses}\label{sec:class_results}

Class-based analyses were performed on the final submissions to gain insights into the composition of the generated image ensembles. The results are reported in \autoref{tab:classwise}. Recall that class information was not made public in the challenge. 

\begin{table}[htb]\label{tab:classwise}
\centering
\caption{Class-based analyses of submissions. Expected class prevalence (\%) from the training data is 10, 40, 40, 10 for the four breast types: fatty, scattered, heterogeneous, and dense, respectively. Class density and coverage is expected to be approximately 1 for all classes.}
\vspace{5pt}
\begin{tabular}{@{}lllll@{}}
\toprule

User & Rank                 & Class prevalence (\%)                         & Class density            & Class coverage \\
\textit{D9}             & \textbf{1} & 10, 41, 39, 10     & 1.0, 1.0, 1.0, 1.0  & 0.9, 1.0, 1.0, 1.0     \\
\textit{K7}            & \textbf{2} & 10, 40, 40, 10     & 1.0, 1.0, 1.0, 1.0  & 0.0, 0.8, 0.9, 0.8     \\
\textit{A8}    & \textbf{3} & 10, 40, 40, 10     & 0.0, 1.0, 1.0, 1.0  & 0.0, 0.8, 0.9, 0.7     \\
\textit{C2}      & \textbf{4} & 3, 30, 55, 12      & 0.0, 0.7, 0.6, 0.2  & 0.0, 0.2, 0.4, 0.1     \\
\textit{V4}              & \textbf{5} & 9, 39, 35, 17      & 0.0, 0.3,  0.5, 0.2 & 0.0, 0.3, 0.3, 0.2     \\
\textit{M3}        & \textbf{6} & 12, 46, 38, 4 & 0.0,         0.2, 0.3, 0.3  & 0.0, 0.1, 0.2, 0.1\\
\textit{J5}     & \textbf{7} & 11, 40, 40, 9      & 0.0, 0.7, 0.8, 0.7  & 0.0, 0.2, 0.7, 0.6    \\
\textit{H1}             & \textbf{8} & 11, 43, 35, 11     & 0.0, 0.4, 0.5, 0.2  & 0.0, 0.3, 0.4, 0.3     \\
\textit{S4}            & \textbf{9} & 14, 42, 40, 4      & 0.0, 0.7, 0.5, 0.5  & 0.0, 0.6, 0.6, 0.4    \\

\bottomrule
\end{tabular}
\vspace{10pt}
\end{table}

For all generated images, class was determined via the four-class classifier described in \autoref{sec:evaluation_strategy} Most submissions demonstrate class prevalence similar to the training data: 10\%, 40\%, 40\%, 10\% for fatty, scattered, heterogeneous, and dense breast types respectively. Only two submissions (\textit{S4}, \textit{M3}) demonstrated instances of mode collapse with class prevalence below 4\% for one class. However, class-wise density, and coverage \cite{naeem2020reliable} computed in the top-2 PC-space of features from all families, were rarely equal or perfect (approximately 1) across classes and submissions. Note that density and coverage are metrics indicative of fidelity and diversity respectively \cite{naeem2020reliable}. Often, at least one class had nearly zero coverage, despite demonstrating ensemble prevalence similar to the training data. Thus, although a generated ensemble may seem to replicate the class prevalence in the training data as determined by a forced-choice classifier, the fidelity of several class-specific features may still be suspect.

\subsection{Results: Analysis of artifacts}\label{sec:further_analysis}

Visual inspection of all generated image ensembles revealed artifacts arising from ligament skeletons, morphology, and texture. Most importantly, some artifacts were not unique to a single submission, but were observed across multiple submissions, suggesting a correspondence to DGM approaches. 

\vspace{10pt}
\begin{figure}[h!bt]
\centering
\includegraphics[width=0.75\linewidth]{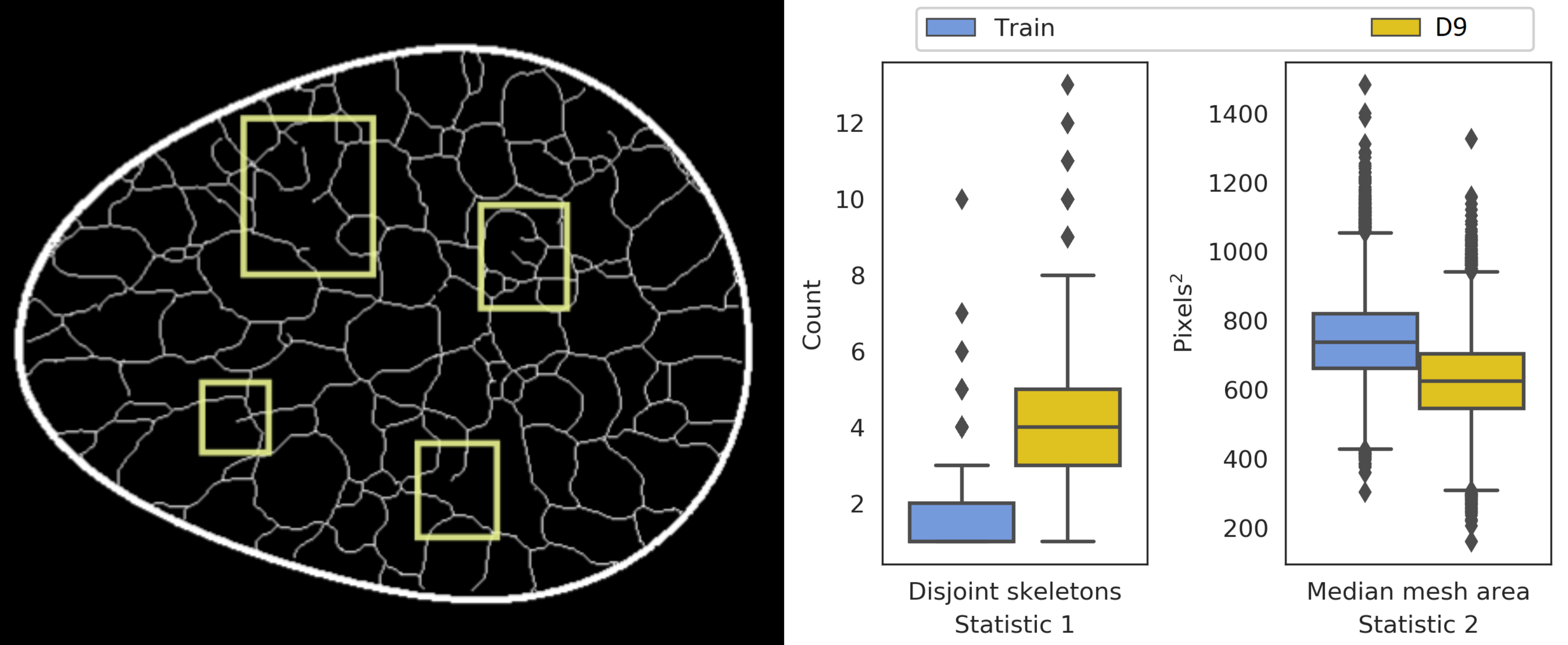} 
\caption{A thresholded sample image (left) from the top-ranked submission (\textit{D9}) demonstrated clear breaks in ligament connections (yellow boxes). These artifacts were also reflected in two statistics: number of disconnected skeletons per-image, and the median area of bounded regions within an image. The boxplots correspond to 10,000 images each, from the training dataset, and the top-ranked submission.}
\label{fig:D9_ligaments}
\end{figure}
\begin{figure}[hbt]
\centering
\includegraphics[width=0.7\linewidth]{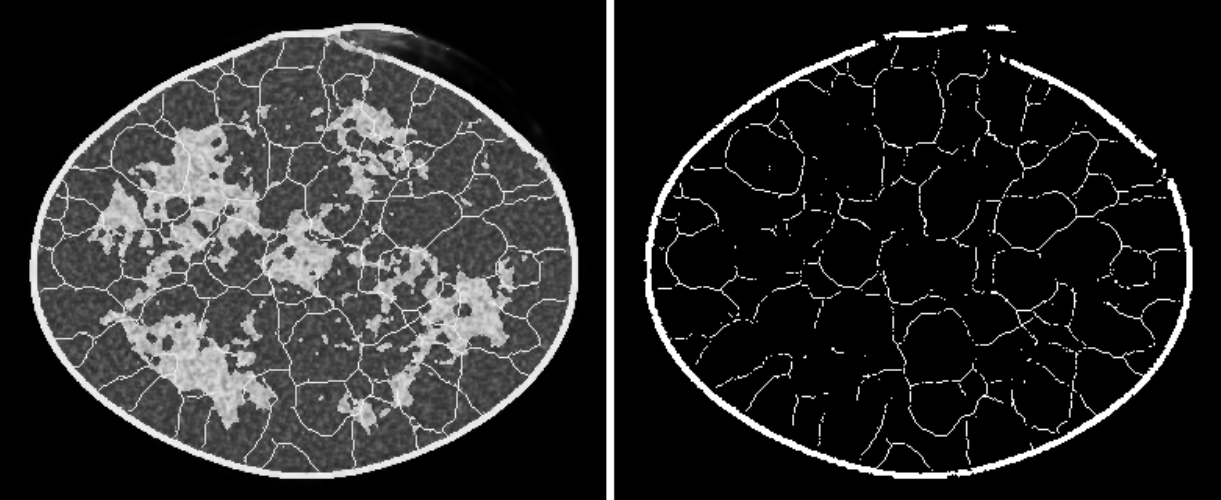} 
\caption{Two kinds of artifacts are demonstrated in a sample image from a submission: (i) broken boundary (left), and (ii) large, smooth breaks in the ligament skeleton (right).}
\label{fig:broken_images}
\end{figure}
\begin{figure}[hbt]
\centering
\includegraphics[width=0.73\linewidth]{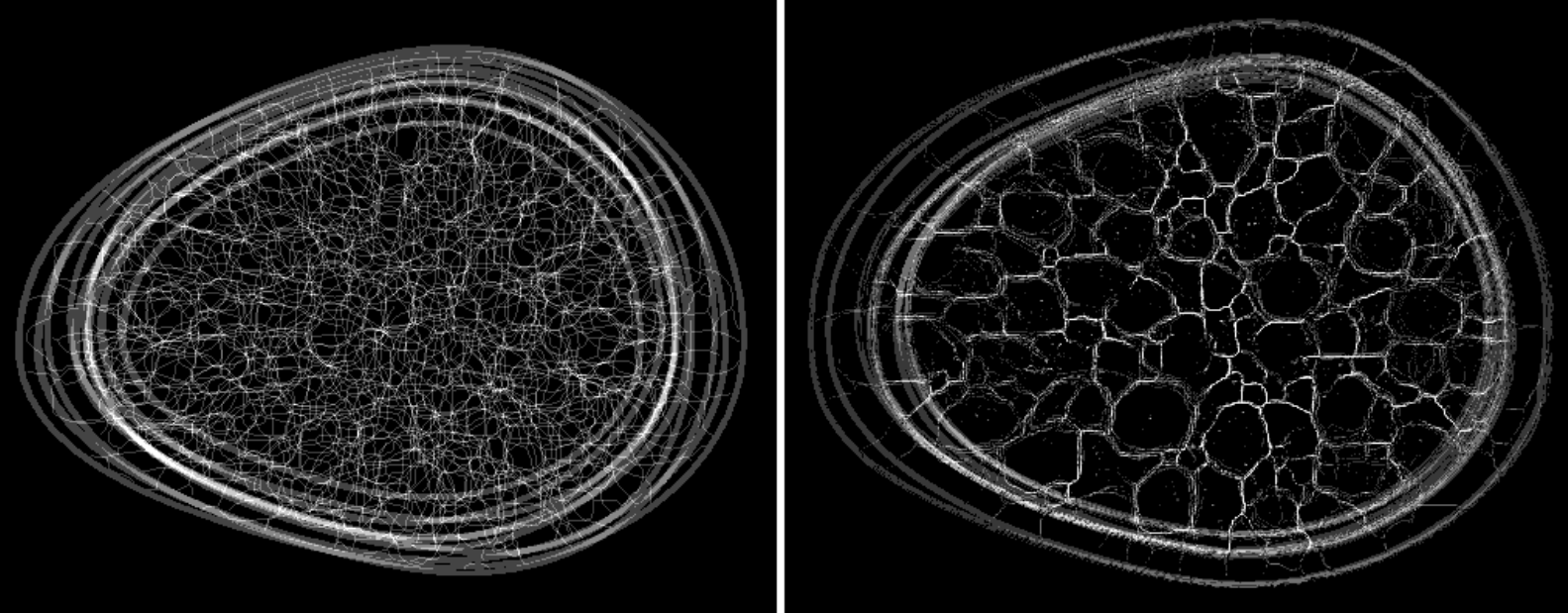} 
\caption{Some submissions demonstrated ligament sticking, that is, nearly constant ligament structure across images. Summation of 10 thresholded images of ligaments from the training data (left) demonstrates the expected randomness in ligament structure across realizations. This variation was clearly absent in similarly processed images (right) from a submission demonstrating this artifact.}
\label{fig:sticky_ligaments}
\end{figure}

All submissions demonstrated artifacts in ligament formation. Three kinds of artifacts were observed in the ligament skeletons: (i) sharp breaks in ligaments, (ii) gradual breaks in ligaments, blending into the background, and (iii) ligament sticking, i.e., constant ligament structure across multiple images. The three artifacts are referred to as ``break", ``blend", and ``stick" respectively in \autoref{tab:errors}. The top-ranked submission demonstrated the first artifact, which was also captured in two statistics from the evaluation framework: (a) the number of disconnected skeletons per-image, and (b) the median region area over all regions in an image (see \autoref{fig:D9_ligaments}). Thus, even the best submission was not perfect. The same artifact was also observed in another submission (rank 6), which also employed a conditional diffusion modeling approach like the top-ranked submission. The second artifact (see \autoref{fig:broken_images}) was observed across all other submissions, indicating that the generation of ligaments was not a trivial task for typical DGMs. Ligament sticking artifacts (see \autoref{fig:sticky_ligaments}) were observed in two of the final nine submissions. Note that this artifact was a genuine effect of the DGM, and not a user-defined feature; ligament sticking was also observed in some of our own experiments with DGMs.

\begin{figure}[h!bt]
\centering
\includegraphics[width=0.45\linewidth]{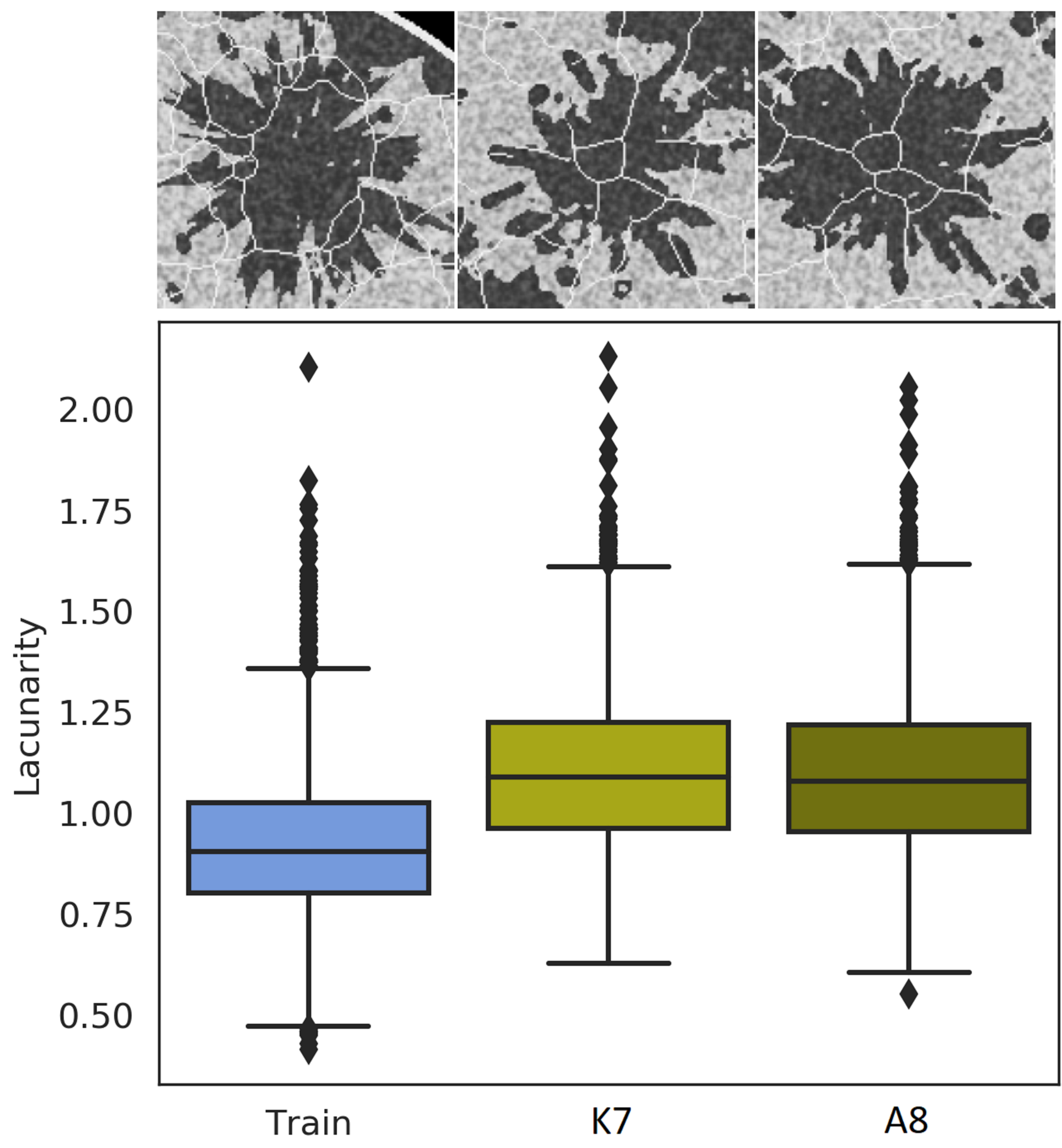} 
\caption{The characteristic ``burst" pattern, typically observed in the heterogeneous breast type, was malformed in several final submissions. Top: An example from the training data (left) shows sharp, splinter-like features, while examples from the second-, and third-ranked submissions (center and right respectively) demonstrate rounded, splatter-like patterns. Bottom: The differences are captured via the lacunarity statistic.}
\label{fig:burst_errors}
\end{figure}

Morphological artifacts of three types were observed across submissions: (i) broken boundaries, (ii) images flipped on the vertical axis, and (iii) malformations in the ``burst"-like features in heterogeneous breast type. The three artifacts are referred to as ``boundary", ``flip", and ``bursts" respectively in \autoref{tab:errors}. Breaks in the boundary (see \autoref{fig:broken_images}), of varying degrees, were observed in three out of nine final submissions. Within these three submissions, the ensemble error rate was observed to be 0.5 to 3\% as determined via a threshold ($<$ 0.9) on the convexity perimeter \cite{wirth2004shape} of the masked breast region. All images from the training dataset were above this threshold. Four other submissions demonstrated instances of images ($<0.1$\% of the ensemble) where the breast region was large enough that it was cut off by the image boundary; these submissions are not flagged for the ``boundary" artifact in \autoref{tab:errors}. The second morphological artifact: flipped images, comprised nearly \emph{half} of the image ensemble for two submissions; this was likely an effect of enabling rotational augmentation during DGM training. The third morphological artifact (``bursts") was specific to the heterogeneous breast type. The characteristic burst-like patterns in this breast type were incorrectly formed in images from five submissions. Examples from submissions ranked 2, and 3, are shown in \autoref{fig:burst_errors}; this artifact was captured in the lacunarity statistic computed on 3000 images classified as heterogeneus breast type, from each submission.

\begin{figure}[h!btp]
\centering
\includegraphics[width=0.45\linewidth]{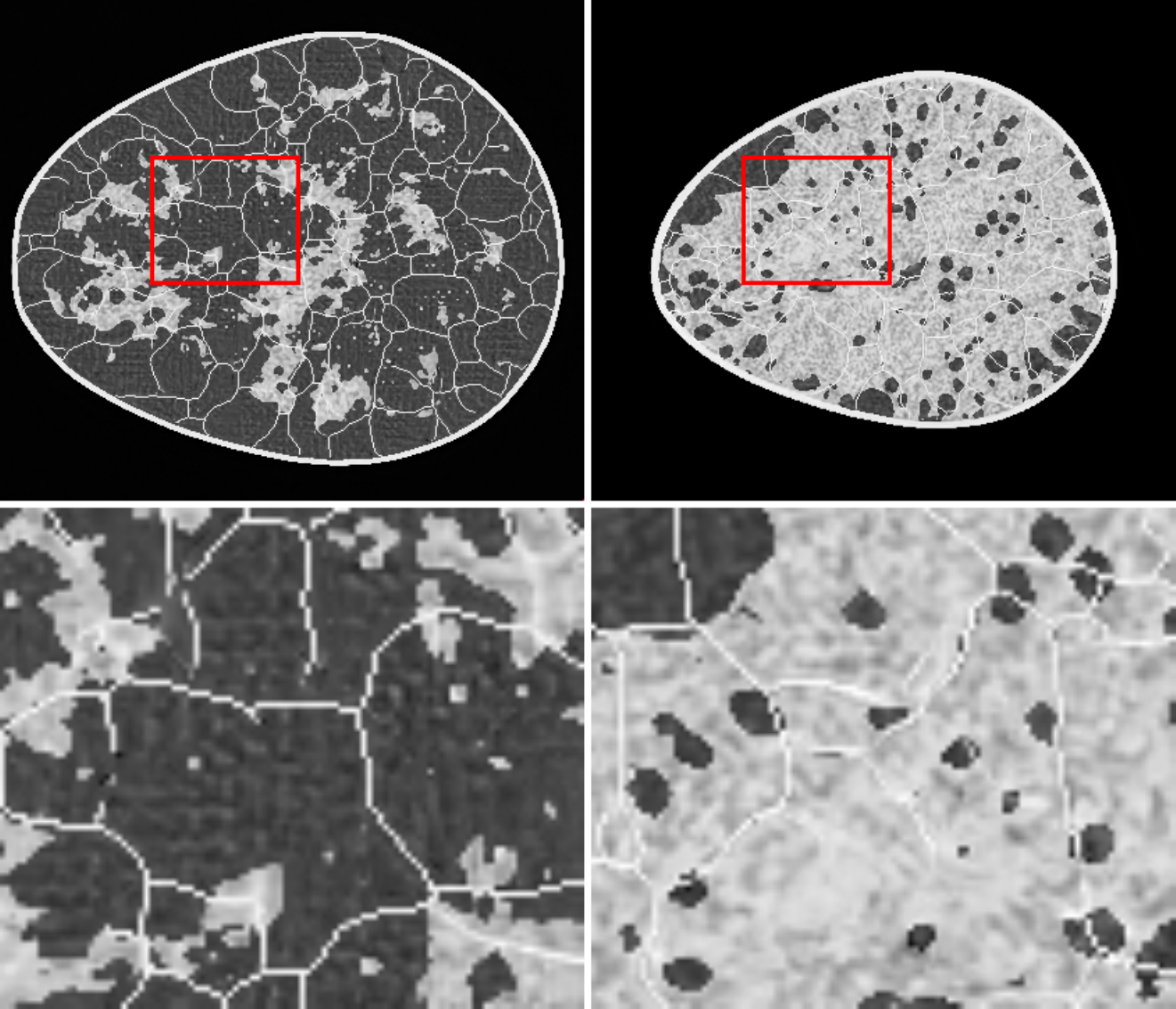} 
\caption{Sample images from two different submissions demonstrating texture artifacts in the foreground region. Gridding, or checkerboard artifacts (left), and ``eddies" (right) were observed.}
\label{fig:texture_errors}
\end{figure}

\begin{figure}[h!btp]
\centering
\includegraphics[width=0.45\linewidth]{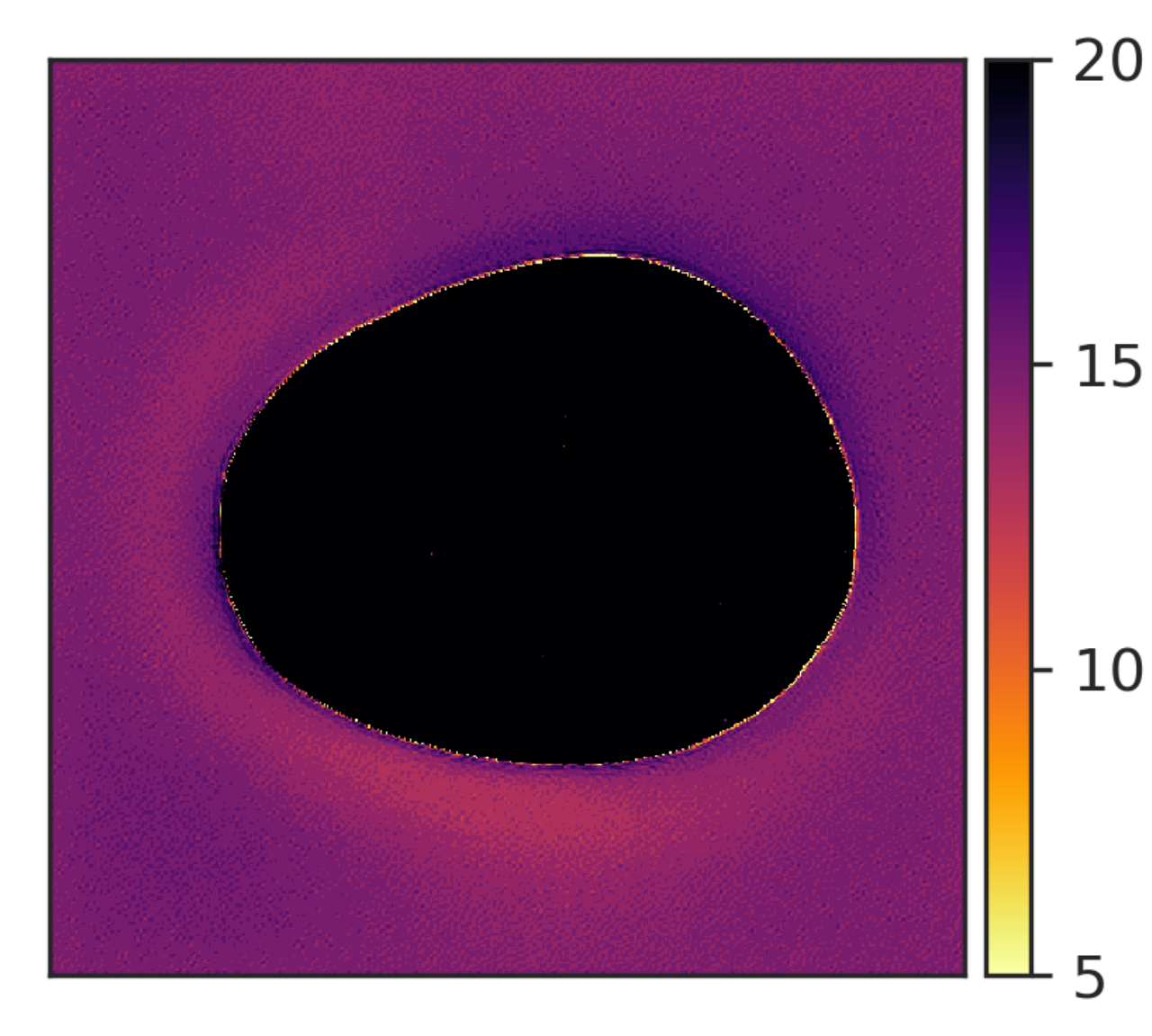} 
\caption{Sample DGM-generated image demonstrating artifacts in the background, which should ideally be constant at zero. Contrast is adjusted for display.}
\label{fig:bad_bg}
\end{figure}

Artifacts in texture were observed in the foreground as well as the background regions. These artifacts are referred to as ``fore", and ``back" respectively in \autoref{tab:errors}. Obvious foreground artifacts were clearly visible in all except the top three submissions. Two such examples are shown in \autoref{fig:texture_errors}. Similarly, background artifacts (see \autoref{fig:bad_bg}), i.e., non-zero background pixels, were found in all except two submissions; however, these were not visually obvious. The mean fraction of per-image, non-zero background ranged from 11 to 62\% over the ensemble for the bottom three submissions for this artifact. However, the ensemble mean of the per-image mean grayscale value over the non-zero background pixels was below 6 for all submissions, and the ensemble standard deviation was at most 18. Thus, the background artifacts were well below the least values in the foreground, and hence, could be eliminated via thresholding. The two submissions that did not have any background artifacts reported employing post-processing techniques on generated images.  

A summary of all artifacts across the final nine submissions is given in \autoref{tab:errors}. A check mark indicates that at least one instance of the artifact was observed in the ensemble. Note that this list of artifacts is not comprehensive, but is indicative of some common artifacts across submissions. Other less visually obvious artifacts may have been captured in the evaluation framework but not described in this section.

\begin{table}[h!btp]\label{tab:errors}
\caption{Overview of artifact types visible in final submissions.}
\vspace{5pt}
\begin{tabular}{@{}llllllllll@{}}
\toprule
\multirow{2}{*}{User}& \multirow{2}{*}{Rank} & \multicolumn{3}{c}{Skeleton} & \multicolumn{3}{c}{Morphology} & \multicolumn{2}{c}{Texture}  \\
& & break & blend & stick & boundary & flip & bursts & fore & back \\
\textit{D9} & 1 & $\checkmark$ & & & & & & &\\
\textit{K7} & 2 & & $\checkmark$ & & & & $\checkmark$ & & $\checkmark$\\
\textit{A8} & 3 & & $\checkmark$ & & & & $\checkmark$ & & $\checkmark$\\
\textit{C2} & 4 & & $\checkmark$ & & $\checkmark$ & $\checkmark$ & $\checkmark$ & $\checkmark$ & $\checkmark$\\
\textit{V4} & 5 & & $\checkmark$ & $\checkmark$ & & $\checkmark$ & & $\checkmark$ & $\checkmark$\\
\textit{M3} & 6 & $\checkmark$ & & & & & $\checkmark$ & & $\checkmark$ \\
\textit{J5} & 7 & & $\checkmark$ & & & & & $\checkmark$ & $\checkmark$\\
\textit{H1} & 8 & & $\checkmark$ & & $\checkmark$ & $\checkmark$& $\checkmark$ & $\checkmark$ & \\
\textit{S4} & 9 & &$\checkmark$ & $\checkmark$ & $\checkmark$ & & $\checkmark$ & $\checkmark$& $\checkmark$ \\

\bottomrule
\end{tabular}
\vspace{10pt}
\end{table}

\begin{figure}[h!btp]
\centering
\includegraphics[width=0.98\linewidth]{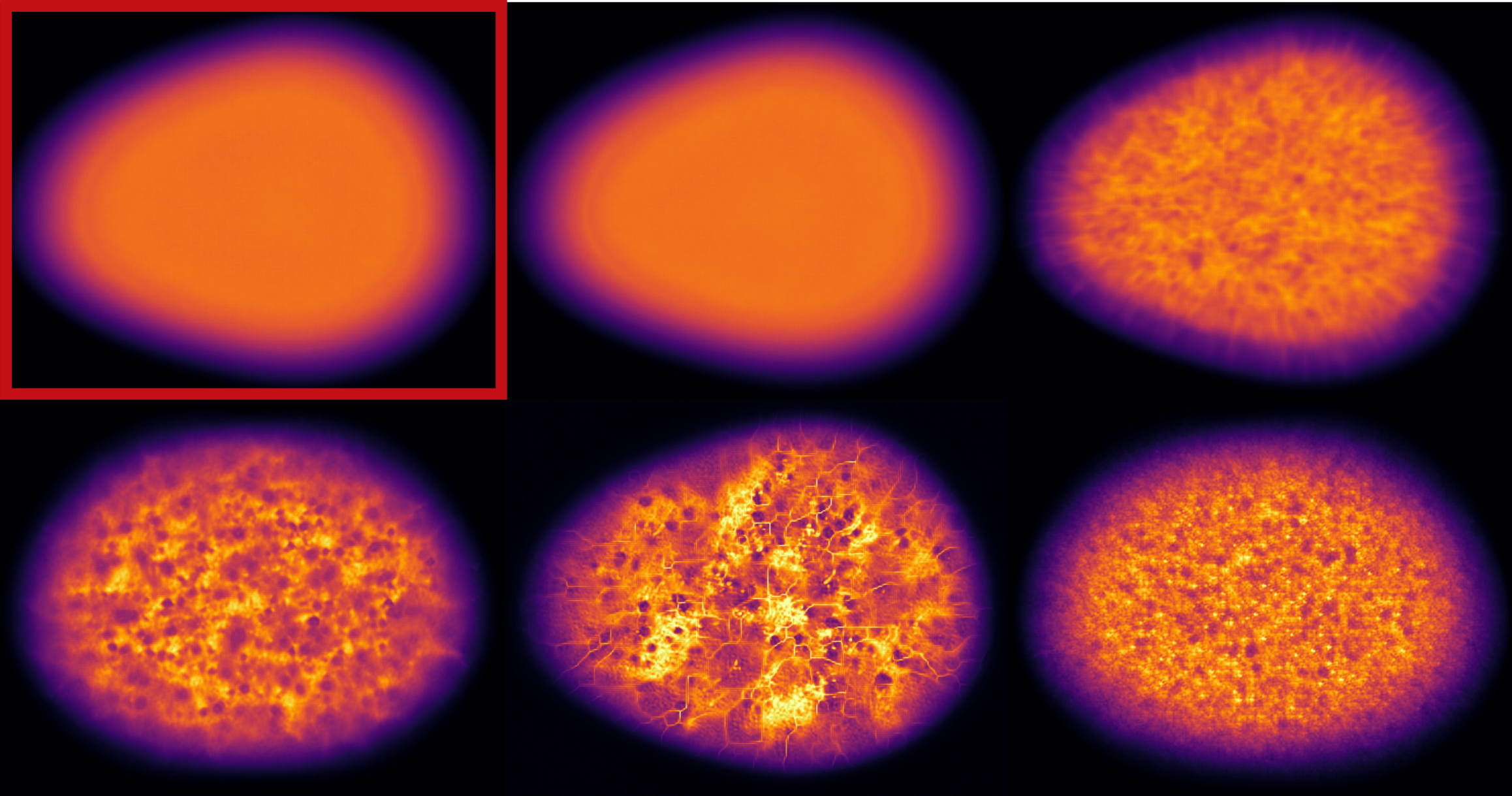} 
\caption{Ensemble mean images over 10000 randomly selected slices from the training data (highlighted in red), and five submissions ranked 1, 2 (row 1, center, and right), 4, 5, and 8 (row 2, left to right), demonstrate positional preference in tissue locations. Distinct structure within the breast region is clearly visible in all except the first-ranked submission, and is indicative of pixel-specific bias in tissue generation.}
\label{fig:sum_montage}
\end{figure}

\begin{figure}[h!btp]
\centering
\includegraphics[width=0.45\linewidth]{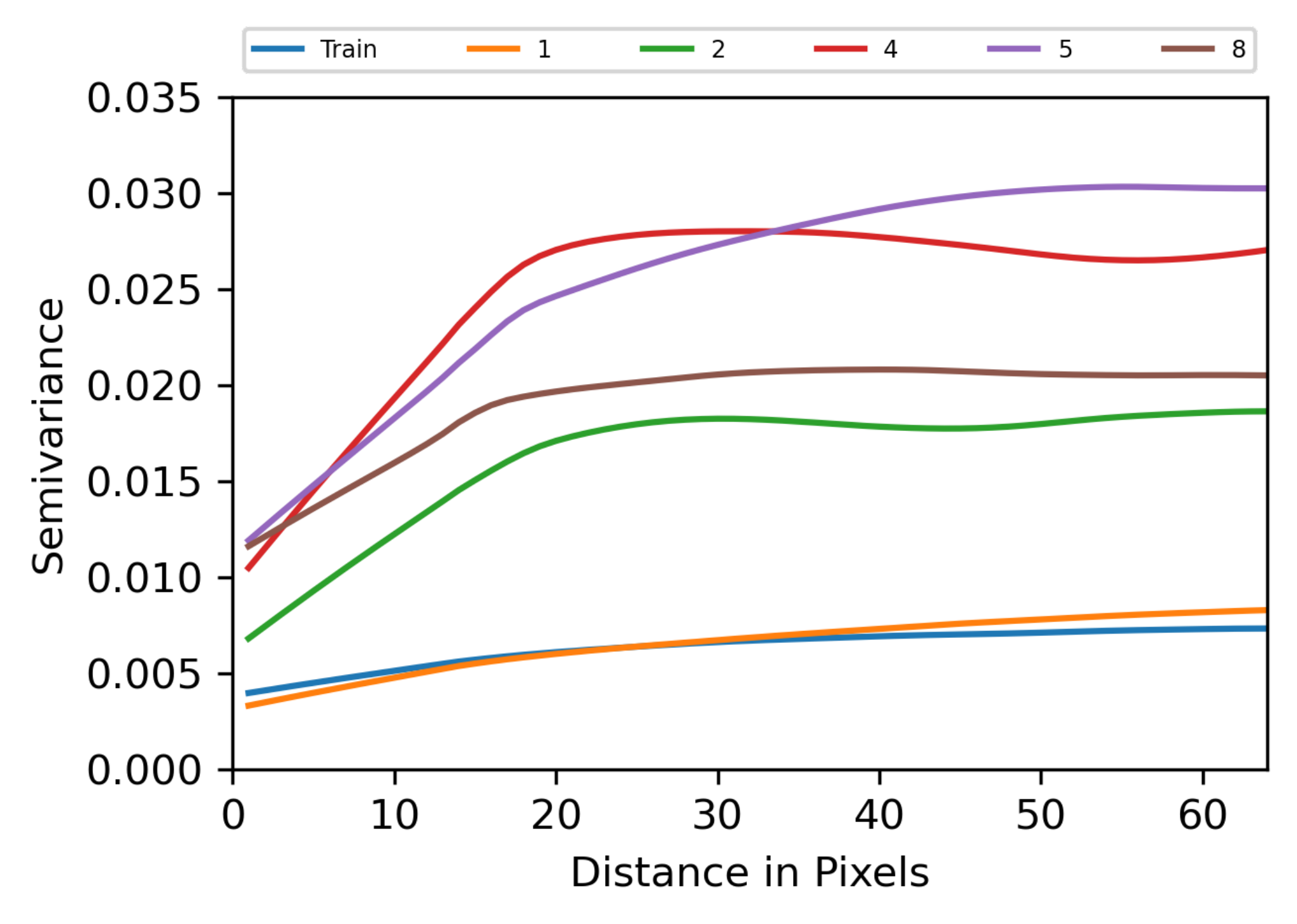} 
\caption{Semivariance of the mean images in \autoref{fig:sum_montage}. The sill of the semivariance determines the inverse of ensemble diversity. Legend indicates overall rank of the submission.}
\label{fig:semivariance}
\end{figure}

Last, we assessed the diversity of an ensemble. From our knowledge of the VICTRE phantom, we intuitively expect ergodicity. In other words, pixel locations should not be strongly tied to tissue identity. Therefore, we computed the semivariance of the mean image obtained from the ensemble (\autoref{fig:sum_montage}). In all but two submissions that were ranked 1 (see \autoref{fig:sum_montage}) and 6 (not shown), pixels at certain positions were preferentially allocated certain breast tissue types, as seen in the corresponding mean images computed over 10,000 images from the respective ensembles (\autoref{fig:sum_montage}). This indicates lower diversity than the training ensemble. Note that the two submissions with ensemble diversity comparable to the training dataset were both conditional latent diffusion models. Because the breast region in the mean images is approximately the same size and centered, the visible differences can be summarized via their semivariances. The long term constancy of the semivariance is indicative of high ensemble diversity. As seen in \autoref{fig:semivariance}, the top-ranked submission is almost as diverse as the training dataset. The  
submissions ranked 2, 4, 5, and 8, clearly demonstrated lower ensemble diversity than the training dataset, and also differed among themselves in terms of the length-scales of pixel correlations.   
\section{Discussion and Conclusion}

It should be clear that objective evaluation and validation of generated images is paramount when these are to be employed in diagnostic medical imaging and analysis pipelines; and, further, that a single number describing the similarity of image ensembles may not directly translate to describe anything meaningful about individual images. Typically, in non-medical demonstrations of DGMs, individual image quality ultimately is assessed subjectively. For example, some humans inspect generated celebrity faces or natural scenes and agree that they sufficiently look like faces or natural scenes but without ever specifying a test statistic used to make that decision. In these examples, we are tacitly and naturally employing domain knowledge of what commonly seen things ought to look like in a binary classification task (whether anything about an image looks wrong). Here, the human is a kind of black-box observer and the \emph{performance} of that observer becomes the objective measure of image quality; if the observer performs excellently at a ``find the fake" task, then the generated images probably are not much like the training data.  

The main challenge with trying to encode hundreds of years worth of domain knowledge about anatomy and physiology into numerical observers is that there isn't a clear, unique mapping between knowledge and tasks. We simply do not always know that performance at one task will accurately indicate performance at another or that accomplishing an arbitrary suite of tasks is informationally equivalent to our domain knowledge. Therefore, the best we can do at this time is to create test statistics and tasks which we know are consistent with our domain knowledge. Thus, the suite of time-tested image-derived statistics that we employed in the challenge analogously is an attempt to numerically say ``what commonly seen anatomy ought to look like" even though we are including only a small fraction of possible descriptors. Additionally, it should be understood that the particular suite of statistics that we employed \emph{likely will not translate well} to other datasets or DGMs evaluations. For example, the use of skeleton statistics was included for the expressed purpose of describing the ligaments seen in the VICTRE phantom. If something skeleton-like is not seen in a different dataset, then the skeleton statistics can become nuisance, or even confounding, statistics. Thus, a key take-away point from the challenge is \emph{not} to test via our particular statistics but, instead, to test, per image, many statistics known to be relevant to the domain under study.

We employed a stochastic object model from which the distributions of various test statistics can be derived without being biased, diminished, or otherwise influenced by an imaging system. Many interesting SOMs could have been employed but we believe that the so-called VICTRE phantom is a particularly good choice for at least three reasons. Foremost, it has been shown to be sufficiently, anatomically realistic for use in virtual image trials \cite{badano2018evaluation}. Second, the population prevalence of various breast types is known and reflected in the training dataset \cite{liberman2002breast}; this is in contrast to many other datasets that comprise only whatever data are available without regard to how many of each relevant kind of thing ought to be present. Third, the two-dimensional cross-sections of the VICTRE object exhibit a wide variety of interpretable structures over a large spatial range. The objects themselves have an egg-like shape on the order of hundreds of pixels across. The distinct fatty and glandular regions transition from hundreds to tens of pixels in size as they punctuate one another. The texture added to glandular regions has a characteristic length on the order of 10 pixels. The ligaments themselves are only a single pixel wide but connected into super-structures spanning 1s, 10s, and 100s of pixels. Taken together, the VICTRE phantom simultaneously exhibits interpretable, testable features at a variety of scales.

We chose the mean grayscale values of tissues to be consistent with the relative separation of x-ray attenuation coefficients such that tissues that would be expected to appear distinct in an x-ray image are distinct in the object. The value of the skin was slightly lowered to make that more distinct from the ligaments. However, unlike for real tissues, we defined tissue-specific intensity distributions (\autoref{eq:intensities}). This was done so that 1) tissues are readily segmentable via thresholding and 2) these distributions are themselves testable features. We observed that none of the entries perfectly reproduced the grayscale intensity distribution of the training data. The upshot is that if exact quantitative values are mission-critical to some use of generated images, then even the DGMs which reproduced spatial statistics well might not be adequate. 

This observation that the distribution of intensity is not necessarily coupled with the arrangement of intensity exposes at least two deficiencies in our evaluation. First, if a DGM reproduced exactly the correct structure but substantially botched the distribution, our naive segmentation would fail. Thus, we might be evaluating a generated ligament framework that looks nothing at all like a genuine ligament framework such that the skeleton statistics significantly differ. A similar consequence would be in that class coverage and density (\autoref{tab:classwise}) appear incorrect as class is defined by the ratio of fatty to glandular tissue; if either segmentation is wrong, so will the ratio be. One way to ameliorate the effects of conflating the variance between classes with the variance within classes is to stratify the generated images by class. However, because, in practice, there usually are no class labels---and a potentially unknown number of classes---in a new dataset, stratification would have to be done via some sort of unsupervised clustering. In our case, because we created the objects, we did have the labels in order to train a classifier. But this exposes a second deficiency: sometimes the generated are so nonsensical that a classifier trained on the training \emph{shouldn't} work well on the generated images. Thus, we can't distinguish a DGM getting one class spot-on while making gibberish off others from mediocre reproduction of each class. This further highlights the need for a straightforward triage of overall performance prior to attempting intra-class, per-image evaluation.

One ready means of ruling out whole-ensembles for further testing is by evaluating the single ensemble mean image (\autoref{fig:sum_montage}). In the case of the VICTRE phantom---which, most pertinently here, is an ergodic model---the mean over numerous images is expected to be roughly constant throughout the bulk of the object with some smooth variation seen near the edge which is due to variation in size of the objects. Semivariance plots \cite{matheron1963principles} can be used to visualize and compare the non-random structure observed in the mean images; in \autoref{fig:semivariance}, the distinction between the top performing model and all others is clear.

Perhaps the most interesting observation is that many of the DGMs made essentially the same kinds of errors (\autoref{tab:errors}). Some of these artifacts most likely are not due to architecture or training but to attempts at data augmentation (e.g., the vertical flipping). The gridding, or checkerboard artifact, shown in \autoref{fig:texture_errors}, has been reported numerous times and for a variety of images \cite{odena2016deconvolution, lee2023impact}, arising as result of network training choices, specifically from the deconvolution layers \cite{odena2016deconvolution}. Other impactful artifacts, however, feasibly are due to architecture. The various challenge entrants trained networks having similar backbones (e.g., GAN or diffusion) to varying degrees while employing a range of hyper-parameters. However, those various networks made many of the same \emph{kind} of mistakes. Another example of this is seen in \autoref{fig:burst_errors}, where the burst pattern is visually less jagged or somehow curvier than the training data. Via visual inspection, we identified the lacunarity as an indicator of this artifact. However, as is seen in the accompanying boxplot, this statistic alone is likely not a robust artifact detector, especially given that the size of the bursts can vary substantially both inter- and intra-class. Bear in mind, too, that we were specifically looking for artifacts. While scrolling through small images on screen, or only seeing a few full-size, the burst pattern sharpness is precisely the sort of thing that can be missed. Thus, an ensemble measure (e.g., the FID) and human visual inspection might both declare a set generated images ``good" even when the details of any one of those don't pass scrutiny. Here, we make no claim that a particular DGM is preferred for all medical images, but instead suggest that anyone training DGMs should be alert to some typical artifact motifs. A reasonable next step would be to design detectors of specific artifacts, across numerous datasets, so that, eventually, training protocols and loss functions can be improved such that the DGMs avoid making the artifacts.

\section*{Acknowledgement}
The authors thank the members of the AAPM Working Group on Grand Challenges (WGGC), in particular, Emily Townley, Karen Drukker, and Lubomir Hadjiyski for their assistance in implementing the DGM-Image challenge. The authors are grateful for the contributions of all the participants of this challenge. The authors also thank Mr. Benjamin Bearce for administering the MedICI platform, and Mr. Muzaffer \"{O}zbey for sharing a trained classifier model that was employed in the analyses in Section IV.B. Varun Kelkar and Dimitrios Gotsis acknowledge funding by appointment to the research participant program at the Center for Devices and Radiological Health administrated by the Oak Ridge Institute for Science and Education through an interagency agreement between the US department of Energy and the US Food and Drug Administration. This work was also funded in part by NIH awards R01EB031585, R01EB034249 and P41EB031772, and by AAPM.

\section*{Conflict of Interest Statement}
The authors have no relevant conflicts of interest to disclose.

\section*{References}
\addcontentsline{toc}{section}{\numberline{}References}
\vspace*{-20mm}









\section*{Figure Captions}
\begin{enumerate}
\item The DGM challenge workflow.

\item Sample images from the training dataset corresponding to four classes: dense (upper left), heterogeneous (upper right), scattered (lower left), and fatty (lower right). Class information was not provided explicitly to the participants.

\item Images generated by the top three approaches alongside the images from the training data.

\item FID and memorization metric scores for the submissions alongside the FID and memorization metric scores of a baseline StyleGAN2 model trained in-house.

\item An image each from three submissions that were ruled out in the first stage of evaluation. The images in the left and center positions correspond to the submissions that did not pass the FID threshold, whereas the rightmost image corresponds to the submission that did not pass both the FID and the memorization thresholds.

\item First two principal components of the features extracted from images from submissions ranked 1, 2, 4, 5 and 8.

\item (Left) A histogram showing the ranking metric values for the 9 submissions that passed the FID-based threshold. (Right) Rank of the submissions with respect to FID plotted against the rank of the submissions with respect to the ranking metric. Note that this plot only shows the submissions that have passed the FID-based threshold.

\item A thresholded sample image (left) from the top-ranked submission (\textit{D9}) demonstrated clear breaks in ligament connections (yellow boxes). These artifacts were also reflected in two statistics: number of disconnected skeletons per-image, and the median area of bounded regions within an image. The boxplots correspond to 10,000 images each, from the training dataset, and the top-ranked submission.

\item Two kinds of artifacts are demonstrated in a sample image from a submission: (i) broken boundary (left), and (ii) large, smooth breaks in the ligament skeleton (right).

\item Some submissions demonstrated ligament sticking, that is, nearly constant ligament structure across images. Summation of 10 thresholded images of ligaments from the training data (left) demonstrates the expected randomness in ligament structure across realizations. This variation was clearly absent in similarly processed images (right) from a submission demonstrating this artifact.

\item The characteristic ``burst" pattern, typically observed in the heterogeneous breast type, was malformed in several final submissions. Top: An example from the training data (left) shows sharp, splinter-like features, while examples from the second-, and third-ranked submissions (center and right respectively) demonstrate rounded, splatter-like patterns. Bottom: The differences are captured via the lacunarity statistic.

\item Sample images from two different submissions demonstrating texture artifacts in the foreground region. Gridding, or checkerboard artifacts (left), and ``eddies" (right) were observed.

\item Sample DGM-generated image demonstrating artifacts in the background, which should ideally be constant at zero. Contrast is adjusted for display.

\item Ensemble mean images over 10000 randomly selected slices from the training data (highlighted in red), and five submissions ranked 1, 2 (row 1, center, and right), 4, 5, and 8 (row 2, left to right), demonstrate positional preference in tissue locations. Distinct structure within the breast region is clearly visible in all except the first-ranked submission, and is indicative of pixel-specific bias in tissue generation.

\item Semivariance of the mean images in Fig. 14. The sill of the semivariance determines the inverse of ensemble diversity. Legend indicates overall rank of the submission.
\end{enumerate}

\end{document}